# Oxygen partial pressure induced dielectric phase evolution in PLD deposited Barium Strontium Titanate thin films

Shashi Pandey[1,2], Suk Hyun Sung[3], Jacob Fernandez[2,4], and John T. Heron[1,2,4]
[1]Department of Materials Science and Engineering, University of Michigan, Ann Arbor, MI, USA, 48109.
[2]Ferroelectronics Laboratory, University of Michigan, Ann Arbor, MI, USA, 48109.
[3]Michigan Center for Materials Characterization, University of Michigan, Ann Arbor, MI, USA, 48109.
[4]Applied Physics Program, University of Michigan, Ann Arbor, MI, USA, 48109.

**Abstract:**

Barium Strontium Titanate ($(Ba_xSr_{1-x})TiO_3$, BST) thin films are amongst the most extensively studied dielectric alloys over the last decade. This is due to the tunability of the dielectric constant and dissipation factor in the underlying ferroelectric and paraelectric states. One important feature of BST is that the competition between the ferroelectric $BaTiO_3$ and the paraelectric $SrTiO_3$ parent phases can be tuned by the A-site alloy concentration, opening a platform for exploring dielectric phase competition. However, the understanding of the interplay between and establishing the control of the dielectric phase from thin film deposition parameters has not been sufficiently reported in the literature. Here, we synthesized a series of $SrRuO_3$/BST/$SrRuO_3$ epitaxial thin film heterostructures by Pulsed Laser Deposition (PLD) from a $(Ba_{0.9}Sr_{0.1})TiO_3$ target under systematic variation of the oxygen partial pressure during the BST deposition. We mapped the dielectric phase of the film to the oxygen growth pressure, film composition, and film crystal structure. A ferroelectric, relaxor-like, or paraelectric phase can be obtained with relatively moderate change in oxygen growth pressure. Scanning transmission electron microscopy (STEM) analysis further indicates that this pressure-dependent phase evolution is accompanied by a structural transition of the BST lattice from tetragonal toward a nearly cubic phase, with the latter corresponding to the relaxor-like state. The control of the dielectric phase opens a pathway to study phase evolution and enable future computing technologies.

**Main text:**

Barium Strontium Titanate ($(Ba_xSr_{1-x})TiO_3$, BST) is one of the most extensively studied dielectric materials over the last few decades due to its high dielectric constant, relatively low dielectric loss, strong dielectric tunability[1-4], and efficient energy-storage capability[5]. As a solid solution of barium titanate ($BaTiO_3$) and strontium titanate ($SrTiO_3$), BST blends the distinct characteristics of its parent compounds: $BaTiO_3$ is ferroelectric at room temperature with a transition temperature $T_C \sim 120$ °C[6], whereas $SrTiO_3$ is paraelectric[7]. Consequently, the Ba/Sr ratio governs the interplay between ferroelectric and paraelectric states in BST, thereby enabling the control of the dielectric properties[8-10] and intermediate dielectric phases.

Several studies have examined dielectric phase evolution in BST as a function of Ba/Sr ratio. Kacimi-Naciri et al. reported that increasing Sr concentration drives bulk BST from the tetragonal ferroelectric phase toward the cubic paraelectric phase[11]. The Curie temperature $T_C$ of bulk BST decreases[12] and the dielectric constant increases[13] with decreasing Ba/Sr ratio. Composition-dependent changes in dielectric behavior and lattice parameters have also been observed in $Ba_xSr_{1-x}TiO_3$ thin films[14]. For PLD-grown BST films, it was found that the dielectric constant increases with deposition temperature[15]. It has been reported that impurity phases form in addition to the BST phase at ~600 °C, and at temperatures above 880 °C, a barium deficiency is observed in the film[16]. Kaura et al. reported a systematic control of composition and epitaxial strain in BST thin films realizing an MPB-like coexistence of cubic, tetragonal, rhombohedral, and orthorhombic phases, enabling high tunability and, at higher Sr content, relaxor-like behavior driven by a diffuse transition and dipole dispersion[17]. Additionally, the influence of laser fluence[18] and laser spot size[19] on the microstructure of thin films has been reported. On the other hand, Gradauskaite et al. demonstrated that the ferroelectric response of thin films is also governed by interfacial electrostatic boundary conditions at the interfaces[20]. By introducing an in-plane-polarized ferroelectric buffer, an artificial flux-closure configuration is created that mitigates the depolarizing field, enabling stable out-of-plane polarization in $BaTiO_3$ from the first unit cell. One of the most sensitive deposition parameters in the complex PLD growth process is the oxygen growth pressure. Lower pressure generally promotes oxygen-vacancy formation while altering the expansion of the laser-generated plasma plume and the kinetic energy of ablated species, thereby influencing the cation stoichiometry, surface morphology, microstructure, growth rate, and crystallographic orientation of oxide thin films[21-23]. Chen et al. reported the effects of oxygen growth pressure on the cationic Ti/Ba ratio, lattice parameter, and microstructure in the thin film of parent phase $BaTiO_3$[24]. However, the detailed, correlated effects of $BaTiO_3$-$SrTiO_3$ phase competition and oxygen growth pressure during thin-film deposition on dielectric-phase stabilization have not been fully or systematically elucidated.

In this work, we demonstrate that a relaxor-like dielectric state can be realized in epitaxial $Ba_xSr_{1-x}TiO_3$ thin films through the coupled control of composition and pulsed-laser-deposition conditions. $SrRuO_3$/BST/$SrRuO_3$ heterostructures were grown from a $(Ba_{0.9}Sr_{0.1})TiO_3$ target while systematically varying the oxygen partial pressure during BST deposition. Correlating the dielectric response with film composition and crystal structure reveals that modest changes in oxygen pressure stabilize ferroelectric, relaxor-like, or paraelectric states. STEM analysis further shows that this phase evolution is accompanied by a structural transformation from a tetragonal toward a nearly cubic lattice, with the latter associated with the relaxor-like response. These findings establish oxygen pressure as an effective control parameter for tailoring dielectric phase stability in epitaxial BST, providing a platform for investigating phase competition and developing tunable materials for emerging computing technologies.

Figure 1(a) shows the workflow used in this study: thin-film growth and optimization, film characterization, device fabrication, and electrical measurements. The $SrRuO_3$/BST/$SrRuO_3$ (SRO/BST/SRO) heterostructures were grown by pulsed laser deposition (PLD) on (110)-oriented $GdScO_3$ (GSO) single-crystalline substrates, varying the oxygen pressure during BST deposition (see Methods). Following synthesis, surface and structural characterization was performed using AFM, XRD, XRR, and STEM. The as-grown heterostructures were then patterned into circular capacitors using photolithography and ion-mill etching. The resulting SRO/BST/SRO capacitors were electrically characterized to investigate their polarization-switching behavior. Figure 1(b) shows cross-sectional annular dark-field (ADF-) STEM images of the SRO/BST/SRO heterostructures grown at oxygen pressures of 150 mTorr (top), 100 mTorr (middle), and 50 mTorr (bottom). The BST and top SRO layers of the 150 mTorr heterostructure exhibit a visibly porous and less dense morphology, whereas the corresponding layers in the 100 and 50 mTorr samples appear comparatively dense and continuous. Figure 1(c) shows typical XRD θ–2θ scans of the SRO/BST/SRO heterostructures with BST films deposited at oxygen pressures ranging from 50 mTorr (green) to 150 mTorr (navy). The top and bottom SRO electrodes are grown at a fixed pressure of 100 mTorr and a slightly lower temperature of 800 °C. The BST 002 peaks gradually shift to higher angles with increasing oxygen pressure, indicating a reduced out-of-plane lattice parameter. The dashed vertical line marks the bulk Ba0.9Sr0.1TiO3 002 peak, at a higher diffraction angle than the films. Energy-dispersive X-ray spectroscopy (EDS) shows that the Sr concentration varies from 7.4% at 50 mTorr to 4.3% at 100 mTorr and 6.1% at 150 mTorr. Compositional variation and epitaxial strain (coherent for 50 and 100 mTorr films) alter the lattice structure and therefore the films' dielectric response.

Figure 1(d) shows typical P–E hysteresis loops measured on lithographically fabricated 25 μm diameter capacitors (see Methods). The P–E loop (10 kHz) of the 100 mTorr capacitor (brick red curve) resembles that of standard ferroelectric $BaTiO_3$, with a saturation polarization of 27 μC/cm² and relatively narrow hysteresis with EC = 47 kV/cm[25]. As the growth pressure is reduced from 100 mTorr, the coercive field initially increases, and the hysteresis loop deviates from the standard ferroelectric-like shape. In particular, the P–E loop of the BST film grown at 50 mTorr is asymmetric, possibly due to oxygen deficiency[26]. When the oxygen growth pressure increases to 125 mTorr, the P–E loop becomes more "S-shaped" or relaxor-like, with reduced saturation polarization and coercive field (magenta). As the pressure increases to 150 mTorr, it becomes more "S-like", while the saturation polarization and coercive field decrease further (navy). Thus, the dielectric behavior of the BST films evolves remarkably across the growth pressure series. To better understand the evolution of the dielectric phase, frequency-dependent P–E loop measurements were performed.

The top panels of Figures 2(a,b) show P–E hysteresis loops for ferroelectric-like and relaxor-like samples (100 mTorr and 150 mTorr samples, respectively). The ferroelectric-like phase (Figure 2(a), top) changes dramatically over the measured frequency range (10–200 kHz), while the relaxor-like phase (Figure 2(b), top) is much less responsive. The bottom panels of Figures 2(a,b) show the frequency dependence of the relative permittivity, calculated as the derivative of the *P–E* loops. Figure 2(c) shows *Ec* vs. frequency and comprehensively illustrates its frequency evolution. The samples with FE-like phase grown at 75–100 mTorr are more susceptible to frequency changes than those with relaxor-like phase grown at 125–150 mTorr. These differences reveal distinct polarization-switching behavior in these systems. In the FE-like 100 mTorr sample, polarization switching is dominated by domain nucleation and domain-wall motion, making *Ec* sensitive to the time available for switching and pinning/depinning. In contrast, the relaxor-like 150 mTorr sample is characterized by more spatially heterogeneous, short-range polar order rather than well-defined long-range ferroelectric domains, resulting in a slimmer *P–E* loop and a weaker frequency dependence[27]. Note that all samples have comparable BST thicknesses of ~115, 95, and 135 nm for 50, 100, and 150 mTorr samples, respectively, and the same capacitor diameter of 25 microns. The 75–100 mTorr samples with FE-like phases and 125–150 mTorr samples with "S-shaped" relaxor-like phases can also be separated into two groups following the frequency dependence of $E_c$, those with higher (lower) $E_c$ together with the larger (smaller) $E_c$ vs frequency slope. For example, the difference in $E_c$ between 20 and 200 kHz for the 100 mTorr sample (navy) is about five times larger than that of the 150 mTorr sample (red).

Figure 3(a) shows the frequency dependence of the remanent-to-saturation polarization ratio ($P_r/P_s$). For all samples, $P_r/P_s$ initially increases with frequency from 20 kHz. However, this increase tends to saturate for the 75–100 mTorr samples, which exhibit predominantly FE-like behavior. In contrast, $P_r/P_s$ continues increasing gradually with frequency for the 125 mTorr sample, suggesting an intermediate response between the FE-like and relaxor-like phase. The 150 mTorr sample exhibits the most pronounced frequency dependence, with $P_r/P_s$ continuing to increase up to 200 kHz, consistent with relaxor-like response. A higher $P_r/P_s$ ratio indicates stronger ferroelectric behavior, greater domain stability, and stronger domain-wall pinning. A lower ratio indicates weaker ferroelectric features and stronger spatially heterogeneous polar disorder, as in relaxor ferroelectrics. Accordingly, the relatively smaller $P_r/P_s$ observed for the 150 mTorr sample compared with the 100 mTorr sample is consistent with the evolution from a predominantly FE-like toward a more relaxor-like phase. Figure 3(b) shows the linear-fit slope of the $E_c$ versus frequency curves in Figure 2(c). The slope is larger for predominantly FE-like samples and smaller for samples with relaxor-like characteristics. This trend indicates a reduced frequency sensitivity of the $P_r/P_s$ response in the relaxor-like regime. Figure 3(c) shows the $P_r/P_s$ ratio measured at 10, 50, and 100 kHz across the pressure series. The ratio remains nearly unchanged between 50 and 100 mTorr, corresponding to the FE-like regime. It decreases as the system evolves toward the relaxor-like state at higher growth pressures, such as 125–150 mTorr.

This pressure-dependent trend occurs at all measured frequencies. Figure 3(d) (left) shows the out-of-plane lattice constant across the pressure series. ADF-STEM measurements reveal that the *c*-value of the BST lattice parameter decreases with increasing growth pressure, consistent with the XRD trend in Figure 3(d) (left). However, the *a*-values change nonlinearly with pressure, leading to the *c/a* ratio versus the growth pressure retaining tetragonality at 50 and 100 mTorr and approaching cubic at 150 mTorr. The reduction in tetragonality at higher growth pressure is consistent with the evolution of the electrical response toward a relaxor-like state. Figure 3(e) shows the area enclosed by the *P–E* hysteresis loop versus oxygen partial pressure at 10, 50, and 100 kHz. The enclosed hysteresis-loop area represents the energy dissipated per polarization cycle, with a larger area corresponding to greater hysteretic energy loss. The inset shows the *P–E* loop of the 100 mTorr sample measured at 10 kHz, with the enclosed area shaded in purple. The hysteresis-loop area remains relatively large for samples grown at 75 and 100 mTorr, which exhibit FE-like behavior. As the system evolves toward the relaxor-like regime at higher growth pressures, the loop area decreases abruptly. At 10 kHz, the enclosed area decreases by approximately a factor of five between the 100 and 150 mTorr samples. This reduction is consistent with the slimmer hysteresis loops and weaker long-range ferroelectric switching associated with the relaxor-like state. Figure 3(f) shows the effective loss tangent, $\tan(\delta_{eff})$ (Methods), measured at 10 kHz versus oxygen growth pressure. $\tan(\delta_{eff})$ peaks at 100 mTorr and drops sharply at higher pressures, following the dielectric phase evolution. The FE-like film at 100 mTorr has the widest hysteresis loops and largest remnant loop area [Fig. 3(e)], consistent with loss dominated by domain switching. In contrast, the relaxor-like films at 125–150 mTorr show slim, S-shaped loops with minimal hysteresis and hence substantially lower loss. In Figures 3(b) to (f), the frequency dependence of polarization, $P_r/P_s$, the out-of-plane c-axis lattice parameter, the area enclosed by the *P–E* hysteresis loop, and effective tangent loss, respectively, all decrease consistently across this pressure interval, providing independent evidence for the phase crossover. This critical pressure regime, therefore, separates the FE-like (red) and relaxor-like (blue) regions.

Next, we measured the temperature dependence of the ferroelectric switching behavior of BST dielectrics, each at three frequencies: 10 kHz (solid curve), 50 kHz (dotted curve), and 100 kHz (dashed curve). For all three samples, the loops gradually become thinner with increasing temperature, indicating a reduction in $E_c$, except on the higher-temperature side of the 150 mTorr sample, which is attributed to increased leakage at elevated temperatures. In contrast, the 100 mTorr sample shows a reduced polarization height as temperature increases, whereas the 50 and 150 mTorr samples show an overall temperature-independent polarization height. Also, the asymmetric bump in the *P–E* loop of the 50 mTorr sample gradually diminished with increasing temperature, and at 200 °C, this feature was almost gone. Figure 4(b) shows the evolution of the $P_r/P_s$ with temperature at different frequencies. Regardless of the sample growth pressure, the change in $P_r/P_s$ looks more pronounced at low frequencies than at higher frequencies. While the $P_r/P_s$ changes non-monotonically with temperature over the measured frequency range for the 50 mTorr sample, particularly for the 100 and 150 mTorr samples, $P_r/P_s$ generally decreases with increasing temperature at each measurement frequency, and increasing the frequency systematically shifts the curves toward higher $P_r/P_s$ values. The temperature dependence indicates reduced polarization stability upon heating, whereas the higher $P_r/P_s$ observed at higher frequencies is consistent with reduced time available for polarization relaxation and back-switching. Figure 4(c) shows the frequency dependence of the $E_c$ at various temperatures. The $E_c$ overall increases linearly with frequency, consistent with the nature of polarization switching. At higher frequencies, the time available for polarization reversal is reduced, thereby increasing the switching voltage. The only difference in trend is that $E_c$ tends to decrease with increasing frequency at 200 °C for the 150 mTorr sample, which is consistent with the broadening of the *P–E* loop and the sudden increase in $P_r/P_s$, as shown in Figures 4(a) and 4(b), respectively. Figure 4(d) shows the temperature dependence of the slope obtained from the linear fits of the $E_c$ versus frequency in Figure 4(c). At room temperature, the 100 mTorr sample with the *FE*-like phase exhibits the strongest frequency dependence of $E_c$, which gradually weakens with increasing temperature. The 150 mTorr sample with the relaxor-like phase exhibits a much weaker frequency dependence that decreases continuously with temperature, becoming negative at 200 °C due to thermally assisted leakage. When the 100 and 150 mTorr curves are compared, overall flat trends are observed with a vertical offset, clearly distinguishing the FE-like and relaxor-like phases. The 50 mTorr curve starts in the intermediate range at room temperature, then spikes at around 120 °C as if it were an FE-like phase; further increasing the temperature returns it to the intermediate phase.

In conclusion, we demonstrate growth-pressure-mediated control of the structural and dielectric phases in epitaxial $SrRuO_3$/BST/$SrRuO_3$ heterostructures grown on $GdScO_3$ (110) substrates by pulsed-laser deposition. XRD and ADF-STEM measurements reveal a systematic decrease in both the out-of-plane lattice parameter and tetragonality (*c/a*) as oxygen pressure increases. This structural evolution accompanies a distinct change in dielectric behavior: the 100 mTorr film exhibits a well-defined FE-like hysteresis response, while the 150 mTorr film displays a slim, S-shaped loop with reduced coercive field and polarization, characteristic of a relaxor-like state. Frequency-dependent measurements of $E_c$ and $P_r/P_s$ further support these phase assignments. Moreover, temperature-dependent measurements from room temperature up to 200 °C show that the distinct frequency scaling of $E_c$ remains discernible throughout the investigated temperature range. Collectively, these results establish a direct correlation among oxygen growth pressure, lattice symmetry, and dielectric phase evolution in nominal $Ba_{0.9}Sr_{0.1}TiO_3$. By leveraging the competition between the ferroelectric $BaTiO_3$ and paraelectric $SrTiO_3$ end members, oxygen pressure offers an effective route to stabilize and tune FE-like and relaxor-like states in epitaxial BST thin films.

**Methods:**

**Target preparation**

10% Sr-doped $BaTiO_3$ (BST) target is prepared by conventional solid-state reaction. The high-purity (>99%) precursors $BaCO_3$, $SrCO_3$, and $TiO_2$ were mixed in the required stoichiometric proportions. The mixture was ball-milled for 3 hours, then calcined at 900 °C for 6 hours. An additional 10 mol% Ba was added to compensate for potential losses from volatilization during target sintering and thin-film synthesis. The calcined powder was granulated in isopropyl alcohol (IPA) and pressed into a 1-inch diameter, 8 mm thick disk at 35 MPa for 15 minutes. The target was then introduced to the first sintering cycle at 1400 °C for 6 hours, with a ramp rate of 6 °C/min. The target was granulated again with IPA into a smooth paste, and the same pressing and sintering steps were repeated to obtain the final dense one inch in diameter pellet used in the film synthesis.

**Thin film synthesis**

SRO/BST/SRO thin film heterostructures were deposited on (110)-oriented $GdScO_3$ (GSO) single-crystalline substrate by using a pulsed laser deposition system with a 248 nm KrF excimer laser. Both SRO and BST films were grown from stoichiometric ceramic targets from Target Materials and from in-house preparation, respectively. During the deposition of the top and bottom SRO layers, the substrate temperatures and oxygen pressures were set to 800 °C and 100 mTorr, respectively, while the BST layers were grown at a slightly elevated temperature of 850 °C, and the oxygen pressure was varied from 25 mTorr to 150 mTorr to synthesize the pressure series samples. The repetition rate of 4 Hz, the laser fluence of ~ 4.3 $J/cm^2$, and the substrate-to-target distance of 6 cm are used throughout the growth. After the top SRO layers are deposited, the chamber is annealed with oxygen pressure of 500 Torr at 800 °C, then cooled to room temperature at 5 °C/min in this oxygen-rich environment.

**X-ray diffraction measurements**

XRD measurements are performed on the SRO/BST/SRO heterostructures on (110)-oriented single crystal GSO substrates using a Rigaku Smart Lab X-ray diffractometer. The 2θ-ω scans were measured with a two-bounce configuration, two-crystal Ge 220 monochromator, and used a Cu Kα radiation source (λ = 1.5406 Å).

**Photolithography and Island Capacitor Fabrication**

A layer of MEGAPOSIT SPR220 3.0 was applied to the heterostructure and spun at 5000 rpm for 30 seconds. The samples were then soft-baked on a hot plate at 95 °C for 3 minutes. The samples were exposed to ultraviolet light for 8.5 seconds through a hard photomask using a Karl Suss MJB3 Mask Aligner. The sample then underwent a post-exposure bake at 95 °C for 3 minutes. The sample was developed with AZ 726 using a 30-30 DP recipe. The developed sample was attached to a 4-inch wafer using Santovac. To define the circular capacitors, the developed samples were etched using a Nanoquest II Ion Mill, which uses Argon as a source gas at 50 sccms. The etch power was 300 W, and the beam voltage was 320 mA. The chamber was held at 5 °°C to prevent the sample from heating during etching, and a SIMS probe was used to monitor the elemental intensities resulting from etching the various layers of the heterostructures, thereby allowing the etch to be precisely stopped at the desired layer height. The sample was tilted at 25 degrees with respect to the ion beam and rotated at 20 rpm to reduce sidewall redeposition and ensure a uniform etch. The samples were then sonicated in acetone and IPA to remove the photoresist, rinsed with DI water, and dried with an $N_2$ gun.

**P-E loop Measurement**

All electrical measurements were performed on circular island capacitors with diameters of 25 µm and 30 µm. The initial *P-E* hysteresis loops are measured on the probe station with one micron tungsten probe tips, using the Precision Multiferroic II from Radiant Technologies. In order to extract the coercive electric field ($E_c$) from the *P-E* loops, first the derivative of the polarization ($P$) with respect to the applied electric field ($E$), i.e., d$P$/d$E$, is calculated separately for the polarization along increasing E and decreasing E and then fitted with the Gaussian function. The center of these two Gaussian fits gives the $E_c^+$ and $E_c^-$, from which $E_C = (|E_c^+| + |E_c^-|)/2$ is calculated. The d$P$/d$E$ is used to calculate the relative permittivity for the BST dielectric.

**Effective Loss Tangent Calculation:**

The effective loss tangent, $\tan(\delta_{eff})$, was calculated from voltage and polarization waveforms measured under 10 kHz triangular excitation. To minimize initial transient effects, the second complete cycle was analyzed, and constant offsets were removed. Fourier analysis was used to extract the fundamental components, and the polarization was resolved into a component in phase with the voltage fundamental ($P_{in}$) and a component lagging it by 90° ($P_{loss}$). The effective loss tangent is defined as $\tan(\delta_{eff}) = P_{loss}/P_{in}$. This quantity describes the large-signal polarization response, including switching and any measured leakage contributions, rather than the small-signal dielectric loss.

**TEM Measurement:**

ADF-STEM measurements were performed on TFS Spectra 300, equipped with Dual-X EDS detector, operated at 300 keV with convergence semi-angle of 21.4 mrad. Electron transparent, thin lamella were prepared with standard gallium focused ion beam (FIB) lift-out process and thinning using Thermo Fisher Scientific (TFS) Helios 5 UC. The samples were thinned down at 30 keV with decreasing current from 0.23 nA and 80 pA, and then with a fine polishing process using 5 keV, 41 pA ion beam.

**Acknowledgements:**

The work is supported by ONR grant N000142412744. S. P. and J. H. acknowledge support from ONR grant N000142412744. Transmission electron microscopy experiments were performed at the Michigan Center for Materials Characterization (MC)$^2$, RRID:SCR_026770) at the University of Michigan. This work was performed in part at the University of Michigan Lurie Nanofabrication Facility and X-ray diffraction facility in Van Vlack Lab.

A meme summarizing this paper can be found at: https://ferroelectronicslab.com/memes/

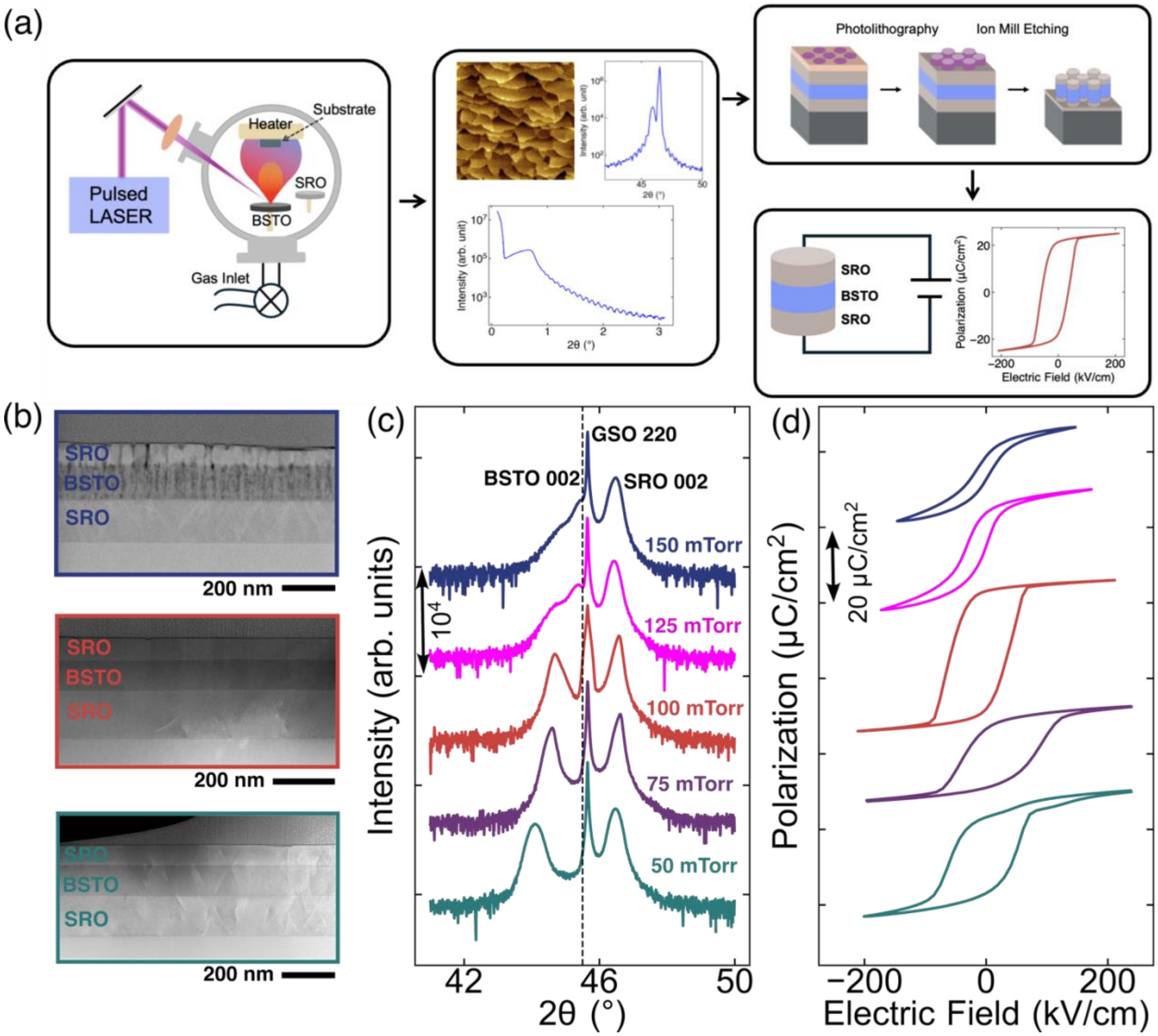

**FIG. 1. Growth, fabrication, and characterization of $Ba_{1-x}Sr_xTiO_3$ (BST) films and the evolution of their dielectric response.** (a) Experimental workflow: pulsed laser deposition of the $SrRuO_3$ (SRO)/BST/SRO heterostructure, characterization by AFM, XRD, and XRR, capacitor fabrication by photolithography and ion milling, and electrical polarization measurement. (b) Cross-sectional TEM images of BST capacitors showing the three-layer structure, with BST grown at oxygen pressures of 150 mTorr (top), 100 mTorr (middle), and 50 mTorr (bottom). (c) XRD 2θ-ω scans of the heterostructures with BST grown at 50–150 mTorr and fixed SRO layers. The BST peak shifts to higher angle, indicating a smaller out-of-plane lattice parameter, as growth pressure increases. (d) P–E loops measured at 10 kHz on 25-µm-diameter capacitors; loop shape and size depend on growth pressure. The spacing between y-ticks is 20 µC/cm². Offsets along the y-axis are added in (c) and (d) for clarity.

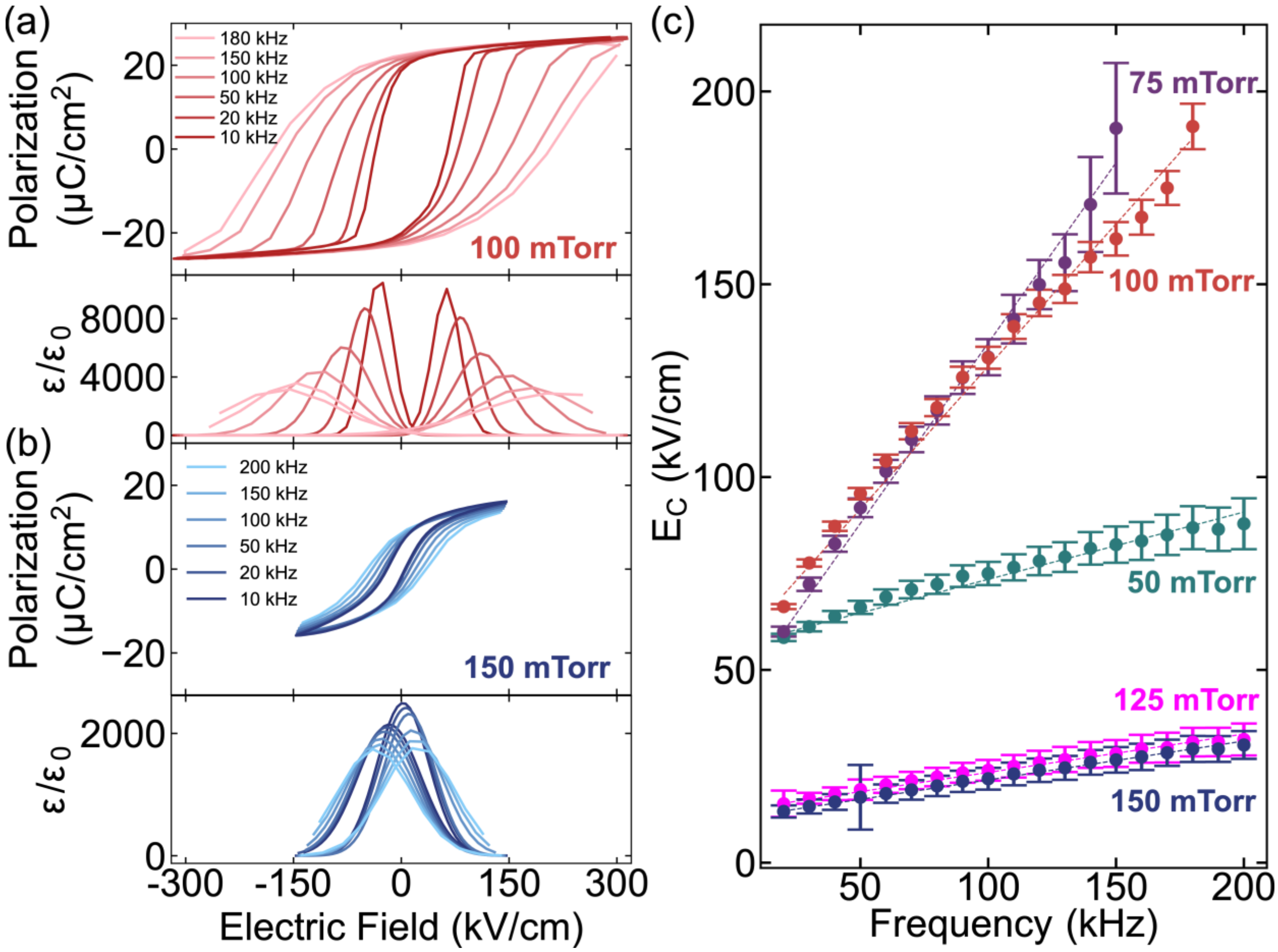

**FIG. 2. Frequency-dependent dielectric behavior of BST films grown at varied oxygen pressure.** (a,b) P-E loops (top) and relative permittivity (bottom) versus frequency for the ferroelectric-like phase at 100 mTorr (a) and the relaxor-like phase at 150 mTorr (b). (c) Coercive field (*Ec*) versus frequency for the full pressure series, with linear fits (dashed). Samples grown at 75–100 mTorr show steeper slopes than those at 125–150 mTorr.

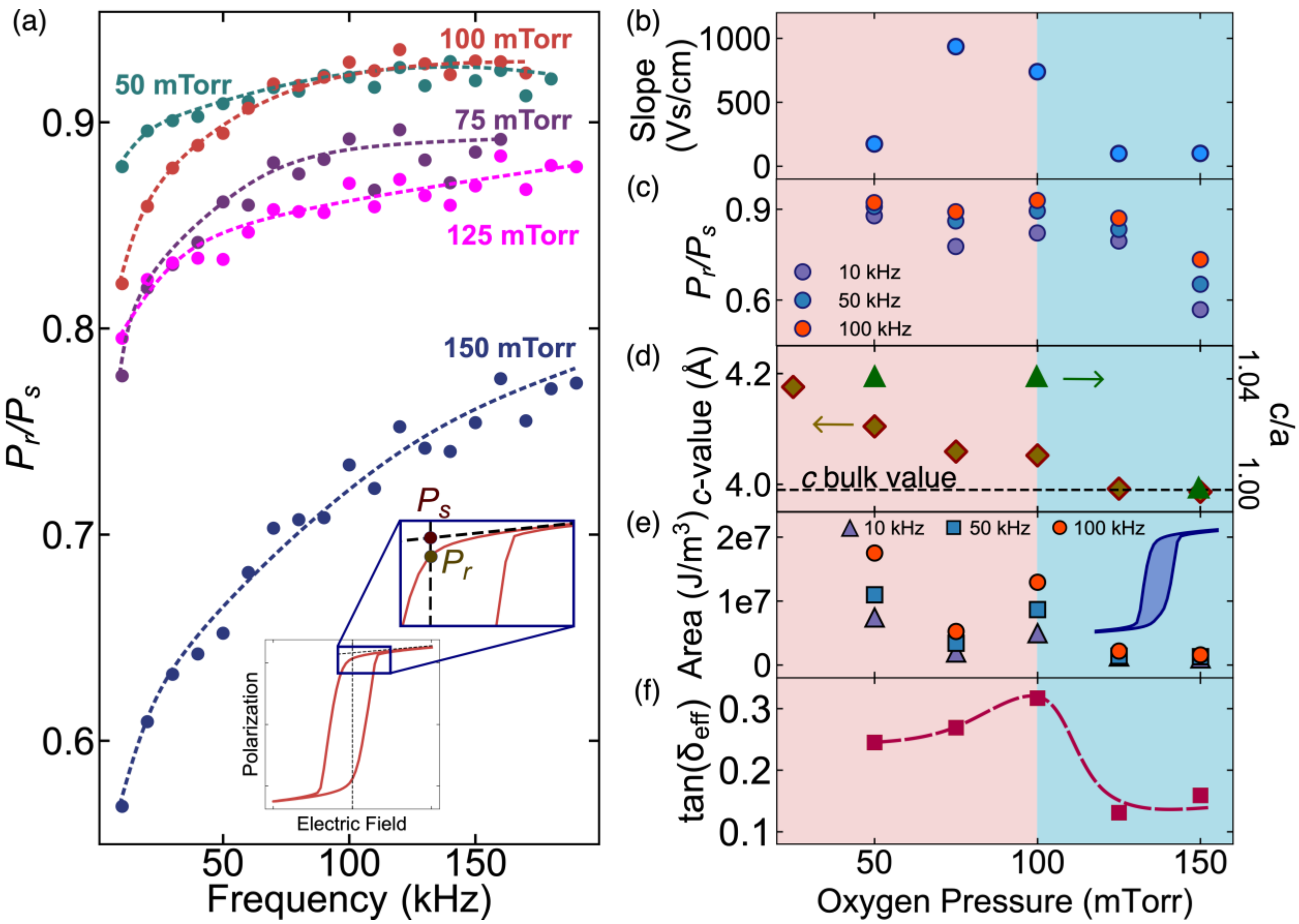

**FIG. 3. Frequency-dependent dielectric behavior of BST films grown at varied oxygen pressure.** (a) $P_r/P_s$ versus frequency for the 50–150 mTorr series; the inset shows how $P_r/P_s$ is determined from a FE loop. (b) Slope of the linear fit of the $E_C$ versus frequency curve in Fig. 2(c), largest for the most ferroelectric samples (75 and 100 mTorr). (c) $P_r/P_s$ at 10, 50, and 100 kHz, peaking at 75–100 mTorr. (d) (left) Out-of-plane lattice constant versus pressure (left) and *c/a* ratio of three samples from STEM (right); The out-of-plane lattice constant decreases with increasing pressure. The horizontal dashed line marks the bulk BST *c*-value. (e) Remnant area of the *P–E* loops at 10, 50, and 100 kHz, and (f) effective loss tangent, tan($\delta_{\mathrm{eff}}$), at 10 kHz versus pressure; tan($\delta_{\mathrm{eff}}$) peaks at 100 mTorr and drops significantly at higher pressures. Dashed lines are guides to the eye.

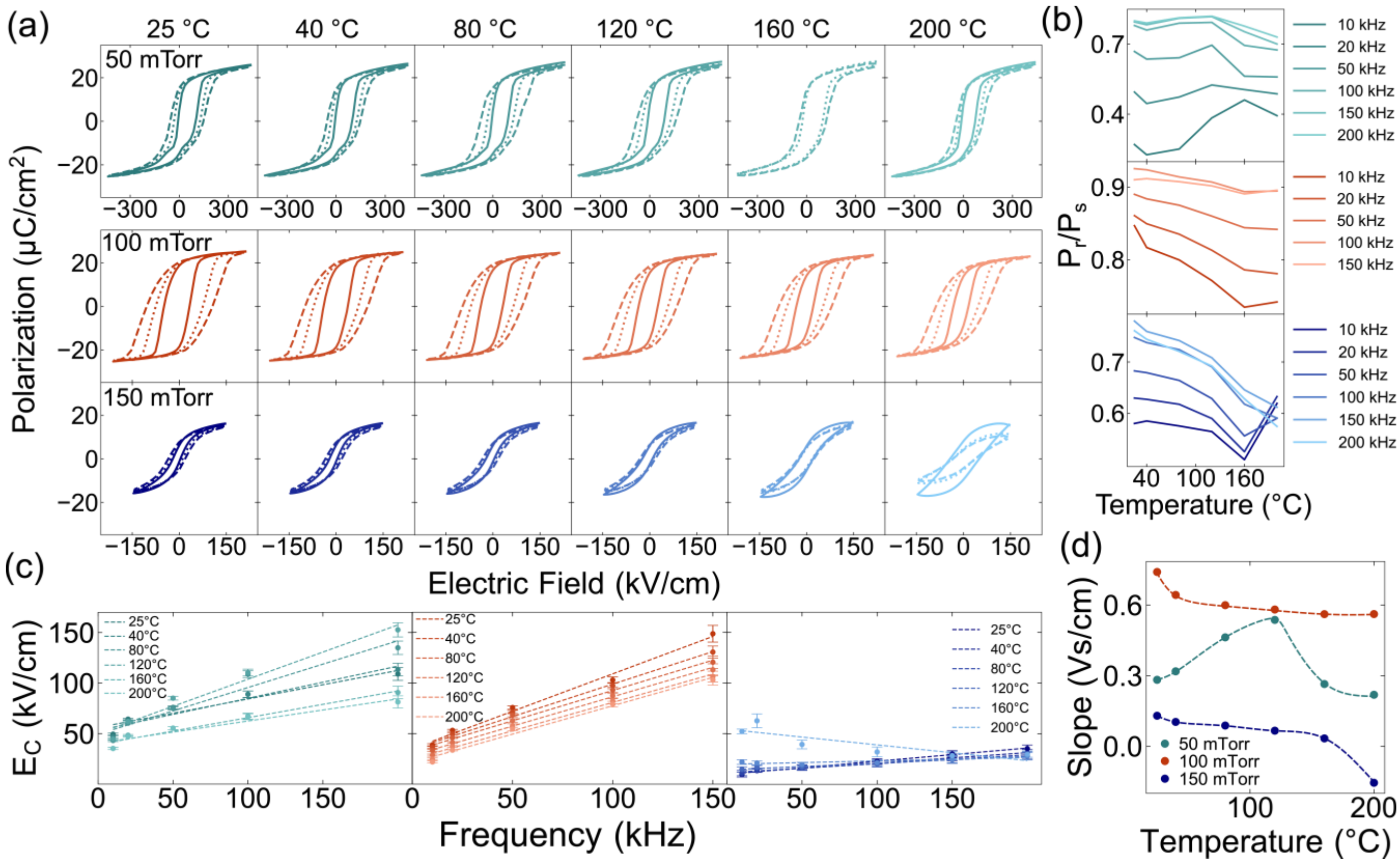

**FIG. 4. Temperature-dependent dielectric behavior of BST films deposited under varied oxygen growth pressure.** (a) Temperature evolution of the *P*-*E* loops. The coercive voltage ($E_c$) and saturation polarization ($P_s$) decrease with temperature, least for the 150 mTorr sample, where the apparent rise in remanent polarization ($P_r$), $P_s$, and $E_c$ at low frequency and high temperature is attributed to the onset of leakage. (b) $P_r/P_s$ versus temperature at different frequencies; the ratio changes more at low frequencies. (c) $E_c$ versus frequency at various temperatures for the 50 (left), 100 (middle), and 150 mTorr (right) samples, with linear fits (dashed). (d) Slopes of the fits in (c) versus temperature. Dashed lines are guides to the eye.